\documentclass[a4paper,10pt]{sigwebnewsletter}

\usepackage{xspace}
\usepackage{url}
\newcommand{\ie}{\emph{i.e.,}\xspace}
\newcommand{\eg}{\emph{e.g.,}\xspace}

\title{Beyond Collaborative Filtering: A Relook at Task Formulation in Recommender Systems  }

\author{AIXIN SUN\\Nanyang Technological University, Singapore}

\newsletterQuarter{June 2024.}
\newsletterYear{ACM DL Version: \url{https://dl.acm.org/doi/10.1145/3663752.3663756}}

\begin{abstract}
Recommender Systems (RecSys) have become indispensable in numerous applications, profoundly influencing our everyday experiences. Despite their practical significance, academic research in RecSys often abstracts the formulation of research tasks from real-world contexts, aiming for a clean problem formulation and more generalizable findings. However, it is observed that there is a lack of collective understanding in RecSys academic research. The root of this issue may lie in the simplification of research task definitions, and an overemphasis on modeling the decision outcomes rather than the decision-making process. That is, we often conceptualize RecSys as the task of predicting missing values in a \textit{static} user-item interaction matrix, rather than predicting a user's decision on the next interaction within a \textit{dynamic, changing}, and \textit{application-specific} context. There exists a mismatch between the inputs accessible to a model and the information available to users during their decision-making process, yet the model is tasked to predict users' decisions. While collaborative filtering is effective in learning general preferences from historical records, it is crucial to also consider the dynamic contextual factors in practical settings. Defining research tasks based on application scenarios using domain-specific datasets may lead to more insightful findings. Accordingly, viable solutions and effective evaluations can emerge for different application scenarios. 
\end{abstract}

\begin{document}

\maketitle


\section{Introduction}
\label{sec:intro}

Recommender System is an attractive research area, evidenced by the increasing number of publications in the past two decades. Based on a prefix search of ``recommend'', about $5000$ publications on RecSys were indexed on DBLP within the year 2023 alone.\footnote{\url{https://dblp.org/search/publ?q=recommend} The query will match publications with titles containing words starting with `recommend', such as `recommender' `recommending' and `recommendation'. The query returns $4943$ publications for the year 2023, and $4460$ for 2022. Note that not all publication titles containing these words belong to RecSys, and not all RecSys papers necessarily include these words in their titles. Nevertheless, based on sampled checking of paper titles in the search results, this prefix search serves as a useful estimation of the number of publications.}  As a reference, only 118 papers, or slightly above 100, were indexed in the year 2002. Given the large and increasing number of publications recently, one might expect the research community to have established a collective understanding of baseline models and evaluation protocols. However, such assumptions may not align with reality.

Before we discuss the research task formulation, we brief the concerns with baseline models and evaluation protocols.
In~\cite{RecBaselines23}, the authors state that ``there are no rigid guidelines that define a comprehensive list of essential baselines''. The authors then created a dataset which ``contains information on 363 baselines used in 903 articles published between 2010 and 2022''. While most popular baselines can be derived from these papers, it is common to receive review comments on the lack of baselines for RecSys paper submissions. That is, the authors and the reviewers do not share a common understanding of a list of must-have baselines.  Even if there were a shared understanding of most performing baselines, there are concerns with the quality of third-party implementation~\cite{thirdPartyImplementation} and hyperparameter tuning~\cite{Winner23}. Thus, reproducibility becomes a concern, and ACM RecSys conference has a recommended list of implementation and evaluation frameworks.\footnote{\url{https://github.com/ACMRecSys/recsys-evaluation-frameworks}} Interestingly, a few large-scale benchmark evaluations show that simple baselines like nearest-neighbor outperform more advanced models (see Table 3 and Section 3.2 in~\cite{EvaluationLandscape24} for a summary and insightful discussion). Probably the most comprehensive evaluation, \cite{McElfreshKV0W22} compares  ``24 algorithms and 100 sets of hyperparameters across 85 datasets and 315 metrics''. All models in the comparison can be the best on some dataset with some metric.  Authors note that ``the algorithms \textit{do not generalize} – the set of algorithms which perform well changes substantially across datasets and across performance metrics". Nevertheless, the simple Item-KNN  is among best performing models, with the highest average ranking position of 2.3 among 20 models (see Table 1 in~\cite{McElfreshKV0W22}). 

In my understanding, various datasets used in RecSys research reflect diverse application scenarios, each necessitating its own set of most effective solutions. However, the simplified task definition abstracts away differences in their practical settings, resulting in model comparability on their prediction accuracies across all datasets. Finding a single model capable of excelling in all application scenarios is challenging. Yet, different forms of nearest neighbors appear to be a common theme in RecSys, to be elaborated further in Section~\ref{sec:view}.

If we rely on the results from these large-scale benchmark evaluations, it seems that no much progress has been made in RecSys, which was a question asked in~\cite{worrying,worrying21tois}. On the other hand, the results of evaluation also heavily depend on the evaluation protocol, in particular, the datasets and the train/test split of a dataset.  A very recent survey~\cite{EvaluationLandscape24} shows that ``the same few (and relatively old) datasets (\ie MovieLens, Amazon review dataset) are used extensively'', and the heavy usage of the MovieLens dataset has also been noted in~\cite{DaisyRec22}. Bauer et al. further comment that ``older datasets may not be good proxies of the user behavior and preferences of today’s users''. In particular, movies rated by a user on MovieLens are those he/she has watched before, hence the dataset cannot simulate the situation of recommending \textit{new} movies to users~\cite{MovieLens23}. Further, the majority of RecSys evaluations do not take global timeline into consideration when splitting a dataset into train and test sets~\cite{FreshLook24,dataLeakage}. As a result, the model under offline evaluation is given access to data records that happen in the future (\eg new items, new users, and also user-item interactions) 
with respect to the time point of the test instance.\footnote{More research is needed to verify whether such data leakage leads to a significant impact on model performances.}  This unrealistic offline evaluation setting may also stem from the simplified formulation of the RecSys task, which overlooks the global timeline.

\section{The Established RecSys Task Formulation }
\label{sec:current}

In Chapter 1 of the \textit{Recommender Systems Handbook}, the core recommendation computation is defined as the prediction of the \textit{utility} (or \textit{evaluation}) of an item for a user~\cite{RSHandbook1Chp22}. The degree of utility/evaluation of user $u$ for item $i$ is modeled as a (real-valued) function $R(u, i)$. Then, ``the fundamental task of a recommender system is to predict the value of $R(u, i)$ over pairs of users and items''. 

We also reference two very recent survey papers at the time of writing. In the survey on modern recommender systems using generative models, \cite{deldjoo2024review} consider ``a setup where only the user-item interactions (\eg `user $A$ clicks item $B$') are available, which is the most general setup studied in RecSys". In the survey on self-supervised learning for recommendation, \cite{selfSSurvey24} formally define the RecSys task with two primary sets:  the set of users $U$ and the set of items $I$. Then, an interaction matrix $U\times I$ is utilized to represent the recorded interactions between users and items, where a value  1 entry means a user has interacted with an item, and 0 otherwise. The definition in~\cite{selfSSurvey24} also includes a notion of auxiliary observed data denoted as $X$; an example is ``a knowledge graph comprising external item attributes". Then a recommendation model aims to estimate the likelihood of a user interacting with an item based on the interaction matrix, and the auxiliary observed data if available. 

The task definitions reviewed above seem to be a common understanding in the RecSys research community. But a few papers  mention the issue of simplification or over-simplification in RecSys research. In Section 7 of the review paper on popularity bias, \cite{PopularityBias23} summarize a few observations, including ``no agreed-upon definition of what represents popularity bias'' in RecSys. The authors further state that ``these observations point to a certain \textit{over-simplification} of the problem and an overly abstract research operationalization, a phenomenon which can also be observed in today’s research on fairness in recommender systems''. In the perspective paper on offline evaluations, \cite{AIM22OfflineChallenge} consider the adoption of  offline evaluation methodologies from experimental practice in Machine Learning and Information Retrieval to RecSys evaluation is a form of simplification. Over-simplification may commonly exists in \textit{research task formulation} in RecSys~\cite{FreshLook24}.  Next, we zoom into the task formulation from three perspectives: user, model, and item.

\section{User, Model, and Item}
\label{sec:view}

The current RecSys task definition mainly involves users, items, and their interactions, in a  \textit{\textbf{static}} view. The task of RecSys is viewed as a task of predicting missing values in an incomplete user-item interaction matrix. Then user-item interaction matrix becomes the key focus of RecSys research.  However, if we examine any specific user-item interaction, it occurs at a particular time point and is the outcome of the user's decision-making. The decision-making can be influenced by various contextual factors. 

\begin{figure}
    \centering
    \includegraphics[trim = 3cm 4.7cm 13cm 1.5cm, clip, width = 0.75\linewidth]{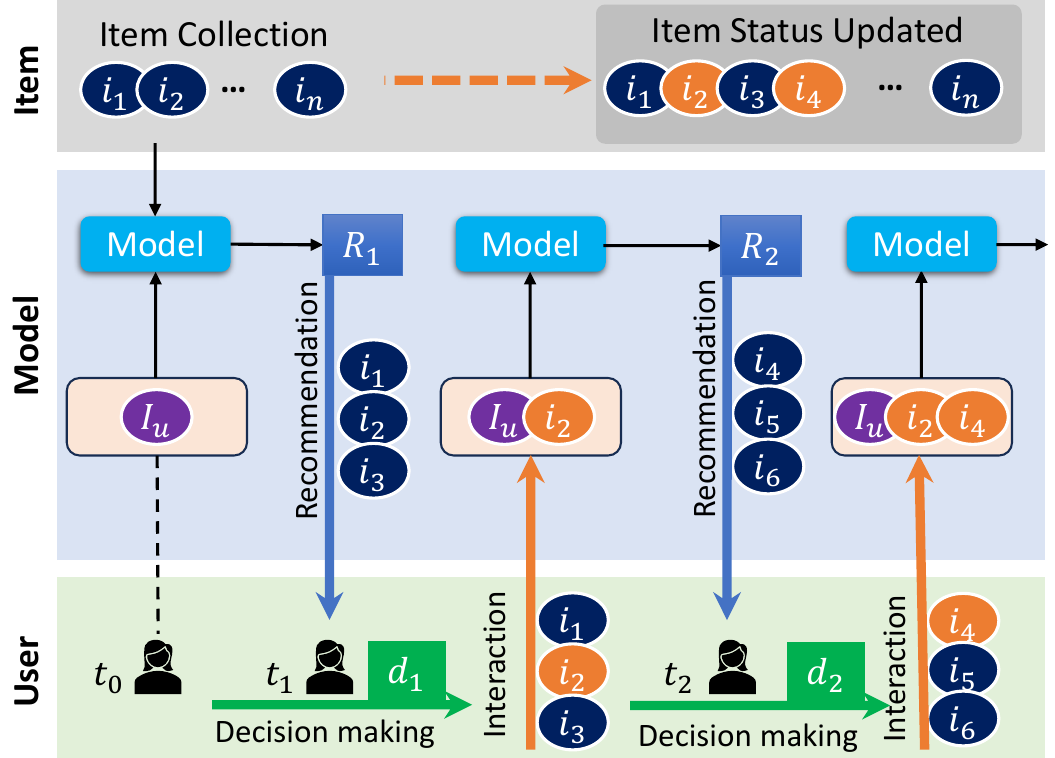}
    \caption{Illustration of two rounds of recommendations made to a user: (i) user triggers a recommender with her past interaction history $I_u$ and receives the first set of recommendations $R1$; and (ii) the user interacts with item $i_2$ after a decision-making process $d_1$, and receives the second set of recommendations $R2$. The user then interacts with $i_4$ after another decision-making $d_2$. Accordingly, the item collection is updated with the two new  interactions. Note that, $R_1$ and $R_2$ are made with different inputs to the model. Best viewed in color.}
    \label{fig:RecProcess}
\end{figure}

We use Figure~\ref{fig:RecProcess} to illustrate interactions between a user and a collection of items, through a recommender \ie a model. We assume that the user is familiar with the recommendation service, and the service provider has the user's past interactions. We also assume that the model is well-trained and its parameters are fixed; its output depends solely on its input.

At time point $t_0$, user $u$ begins interacting with the recommendation service by opening a mobile app or a website, such as YouTube for video viewing or Amazon for products. Based on the set of items that $u$ has interacted with before $t_0$, denoted by $I_u$, the model makes recommendations $R_1=\{i_1, i_2, i_3\}$ from a pool of candidate items. Upon receiving the recommendations $R_1$ at time $t_1$, $u$ considers these three items and chooses to interact with $i_2$.  Accordingly, right after time $t_1$, the interaction records available to the model would be $I_u \cup \{i_2\}$. With the current input of $I_u \cup \{i_2\}$, the model makes the next round of recommendation $R_2=\{i_4, i_5, i_6\}$  at time $t_2$. Upon further consideration,  the user selects $i_4$ for interaction. Consequently, for the subsequent round of recommendations, the model assimilates additional knowledge from $I_u \cup \{i_2, i_4\}$ to make more accurate predictions regarding the user's current interests within the current interaction session. 

Here, we assume that the two recommendations, $R_1$ and $R_2$, occur consecutively within a single session of interactions. It is important to note that in this illustration, we distinguish between the two newly available interactions $\{i_2, i_4\}$ and the past historical interactions  $I_u$. This distinction is made because interactions to $\{i_2, i_4\}$ just occurred, while $I_u$ may have occurred much earlier, with respect to the current session.

From the \textbf{user}'s perspective, at time $t_0$, upon opening the recommendation service, the user expects the recommender to accurately predict her \textit{latent needs} on information, services, or products. Upon receiving recommendations $R_1$ at $t_1$, the decision to interact with $i_2$ is the outcome of a \textit{decision process}, represented by $d_1$ in the figure. This decision process may consider various factors, such as attributes of the recommended items, the user's current location, time of day, ongoing activities, and even the user's mood. For example, users may choose to watch different types of videos on YouTube depending on whether they are feeling happy or sad. The interaction with $i_4$ is the outcome of another decision process.

From the \textbf{model}'s perspective, the two sets of recommendations are generated by using different user-side inputs: $I_u$ for $R_1$, and $I_u \cup \{i_2\}$ for $R_2$, respectively. If a subsequent recommendation is to be made, the user-side input will be $I_u \cup \{i_2, i_4\}$. Moreover, the user's decisions to interact with $i_2$ from $R_1$ and $i_4$ from $R_2$ may strongly suggest that the user, at \textit{the current moment}, is interested in items similar to or related to $i_2$ and $i_4$, yet confined by the overall preference demonstrated through $I_u$. Taking videos as example items, similar videos to $i_2$ and  $i_4$ include videos in the same genre, uploaded by the same content creator, or featuring the same actors, and yet not too distinct from those viewed in the past. The relationships between items may also be established through various means, \eg by content similarity, by collaborative filtering, or other forms of knowledge. In many cases, $I_u$ serves as a valuable resource for understanding a user's general and enduring preferences, gleaned from past interactions. In the ongoing interaction session, the identification of $\{i_2, i_4\}$ reveals the user's current interests, prompting recommendations of similar or related items. This could explain why item-KNN continues to perform well in many evaluations. 

From the \textbf{item}'s perspective, as depicted in the upper portion of Figure~\ref{fig:RecProcess}, after the two interactions from user $u$, both $i_2$ and $i_4$ each receive an additional interaction.\footnote{While user-item interactions may be recorded separately in a system, some aggregate attributes, such as the number of views, likes, and orders/sales from all users, are updated in real time. } If we consider the number of interactions an item receives as its \textit{popularity} attribute, then the attributes of both items change. Given that a typical recommender system serves a substantial number of users concurrently, such attribute changes can significantly impact a large number of items within a short period. In extreme cases, a popular video can attract thousands or even millions of views within a day or two. 

In short, from three perspectives of user, model, and item, a recommender system should be viewed in a \textit{\textbf{dynamic}} setting, instead of a prediction of missing values in a static user-item interaction matrix. However, the dynamic nature of RecSys is largely overlooked in academic research, as the time dimension is often omitted from RecSys task definitions. The ignorance of the global timeline in the task formulation is also the root of data leakage in offline evaluation.  More importantly, current task definitions do not sufficiently focus on the \textit{decision-making} process~\cite{DecisionMakingEC22,Jameson2022humanfactor}. 

\section{Recommenders are Task-Specific}
\label{sec:taskspecific}
The recommendations generated by a model are influenced by the information it gathers from its inputs. These inputs vary significantly depending on the application scenario. Let us consider food recommendation as an example application. When users open a food delivery app, they place their orders and await the arrival of their chosen dishes. Effective recommendations streamline the ordering process, potentially reducing browsing time and allowing users to receive their food earlier.\footnote{Our focus here is on food recommendation within the food delivery app. We do not consider the problem of selecting the best route for delivery.}

Following the current practice, a typical task formulation is: Given a set of users $U$, a set of food items $I$, and a user-item interaction matrix $U\times I$ storing the past orders of all users, the task is to predict the likelihood of users place orders on the food items. Again, this is an oversimplified problem definition. Taking the application scenario into account, it is crucial to recognize the diverse food preferences people have for breakfast, lunch, and dinner. Therefore, considering the time of day in food recommendations becomes essential. Additionally, since the order is for delivery, the anticipated delivery time significantly impacts user experience. There is a notable correlation between delivery time and the distance between the user and the food store fulfilling the order. The ordering time and delivery address are both available to the app and the recommendation model. As a piece of domain-specific knowledge, there is a strong presence of repeat patterns in food consumption  \ie users often repeatedly place orders from the same store~\cite{RepeatFood24}. 

The task formulation shall then be revised. We have a set of users $U$, a collection of food items $I$, and a collection of past transactions as user-item interactions $U\times I$ with auxiliary detailed information such as order timestamps and delivery details. When user $u$ triggers a food delivery recommendation request at time $t$, the application is tasked to provide recommendations by considering the temporal context, the delivery address, all previous orders ($U\times I$), and the user's  purchase history ($I_u$).

Note that $I_u$ is part of $U \times I$, yet it is necessary to separately consider $I_u$ due to the recurrent pattern; users frequently reorder food they have previously tried. The primary distinction lies in the item search space: for repeat orders, we recommend items from within $I_u$, whereas, for first-time or exploration orders, the search space is $I - I_u$; and $|I - I_u|\gg |I_u|$ in terms of cardinality. Due to the significant difference in the search space, separate recommendation models can be designed for repeat and exploration orders respectively~\cite{RepeatFood24}.

In this case study, the task formulation has evolved beyond simply predicting missing values in a basic and static user-item interaction matrix, now incorporating additional spatial and temporal dimensions. The strong presence of domain-specific knowledge on repeated consumption significantly influences the design of recommendation models, as this knowledge directly affects the item search space. Such additional information in the input may or may not affect the recommendation modeling, but these factors may heavily affect the user's decision-making from the user's perspective. Similar recommendation scenarios include hotel recommendations and restaurant recommendations. In both cases, recommendations are more meaningful when a user plans to visit an unfamiliar place. The destination to be visited should be considered as a known input to the recommendation model, \eg booking a hotel for  a conference two months later in a different city.  Similarly, if a user explicitly expects songs from a specific singer or movies featuring a particular actor/actress or genre, these preferences become part of the input. It is important to note that while related, these additional inputs differ from the side information widely studied in recommendation systems~\cite{sideInfoSurvey19}. Side information, such as knowledge graphs about items, may only be available to the RecSys model and remain transparent to users. Furthermore, the exploration of side information has primarily focused on enhancing model accuracy or addressing issues like cold start or cross-domain settings, which users are typically unaware of, and are not factors in users' decision-making.

Based on the above discussion, we may consider a recommender taking the following inputs: $\langle X, u, I_u, I_c, I, U \times I \rangle$. Among them, $X$ represents task-specific contextual inputs such as time and location. Note that, the $X$ here refers to those contextual factors that are available to and/or accessible by the users or even the explicit input from users \eg movie genre. These factors are part of the user's decision-making consideration and are not the kind of auxiliary knowledge only known to the model. Among the remaining inputs, $u$ denotes the user who initiates the recommendation service, $I_u$ consists of the user's past interacted items, and $I_c$ comprises the newly interacted items in the current session (\eg $\{i_2, i_4\}$ in Figure~\ref{fig:RecProcess}).  $I_c$ is empty at the start of the current session, and changes along with the increasing availability of new interactions. $I$ refers to all candidate items available for recommendation.  With the increasing available interactions, some of the interaction-related attributes of items in $I$ are dynamically updated \eg the number of interactions an item receives. $U \times I$ represents historical user-item interactions before the current session.\footnote{Strictly speaking, $U \times I$ is also updated along with new interactions in the current session. Typically the number of newly available interactions within a short time is very small compared to all historical interactions.}  The task of a recommender is to generate recommendations for user $u$ under the current decision-making context.  For comparison, the inputs considered in common RecSys task formulations are $\langle u, I, U\times I\rangle$.

\section{The Mismatching Datasets}
\label{sec:datasets}

We have outlined the inputs that a recommender system should consider, primarily to highlight the dynamic nature of RecSys by emphasizing the context of decision-making.  Note that, the consideration of both user's  general preferences and the current contexts is not new at all. \cite{WideDeep16} consider not only user (static) features like country, language, and age, but also contextual features like device,  hour of the
day, day of the week for app recommendation. \cite{AliDIN18} also consider context features for click-through rate prediction.
A ``User Instant Interest'' modeling layer is part of the solution proposed in~\cite{InstantWSDM24} to model the user's current interest following the user's behavior \ie clicking an item that is referred to as a trigger item. The assumption is that ``the clicked trigger item explicitly represents the user’s instant interests". \cite{Www16List} define the list recommendation problem: Given a specific user $u$ at time $t$, the goal is to produce an ordered personalized list of $K$ that maximizes the probability that $u$ will click on an item
from the list. Here, the current time $t$ is part of the problem definition. The proposed solution also considers contextual features like the number of times an item has been presented to a user and the time of day. Without surprise, all aforementioned papers are from industry.

Eventually, in a practical setting, the recommendation is a ranking problem with at least two forms of latent needs of information/service/product, learned from  (i) the relatively static user/item features and historical interactions, and (ii) the \textit{current and dynamic} interaction process, respectively. The learned preferences then serve as \textit{implicit queries} for item ranking or re-ranking~\cite{Reranking22Survey}.

However, in academic research, accessing an online recommendation platform is not feasible in most cases. The \textit{comprehension of the current context}  relies on two factors. The first factor is the information available in an offline dataset. The widely used datasets like MovieLens and Amazon reviews only record the outcomes of the decision-making process, but not the context of decision-making \eg under what consideration and/or among which options, a user decides to watch a movie or buy a product. The second factor is the way a dataset is used, \eg whether the user-item interactions are arranged in chronological order following the global timeline, and how a model is trained and evaluated on the dataset.

\begin{figure}
    \centering
    \includegraphics[trim = 4cm 7.8cm 16.2cm 2.5cm, clip, width = 0.65\linewidth]{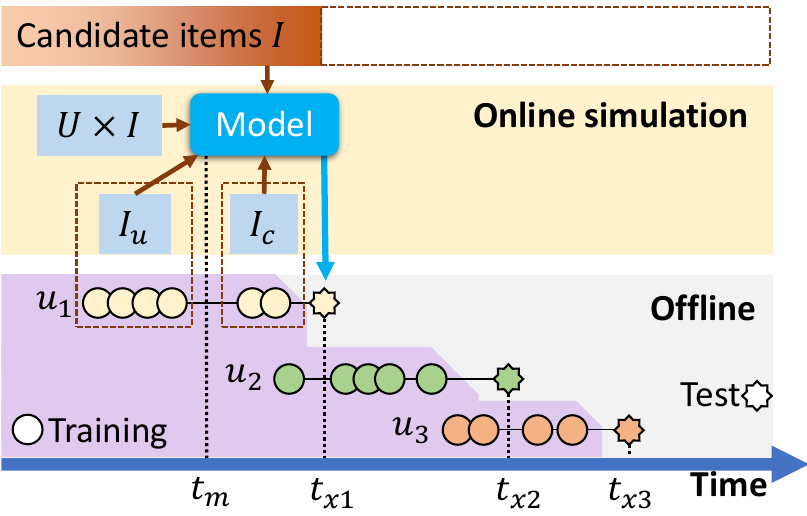}
    \caption{An illustration of train/test instances using leave-one-out data split with interactions by three example users. All interactions are arranged in chronological order following the global timeline. The last interaction of each user is the test instance, represented by a squared octagonal star. Circles are training interactions. The lower half of the figure shows the train/test instances in a typical offline evaluation. The upper half of the figure shows an ideal simulation of a model trained/updated at time $t_m$, for predicting $u_1$'s test instance. Best viewed in color.  }
    \label{fig:context}
\end{figure}

We further elaborate on the second factor through Figure~\ref{fig:context}. The figure shows an illustration of train/test instances using leave-one-out data split with interactions by three example users. Here, all interactions are arranged in chronological order following the global timeline. The last interaction of each user is the test instance, represented by a squared octagonal star, and the circles represent training instances. Following a typical offline evaluation, illustrated in the lower half of the figure, a model is trained by using all training instances, and then evaluated on all test instances. Take user $u_1$ as an example. Many training interactions in the dataset occurred after $u_1$'s test instance \ie time $t_{x1}$. Then the model predicts $u_1$'s test instance, with future training data that occurred after $t_{x1}$. This is a data leakage issue discussed and evaluated in our earlier work~\cite{dataLeakage,FreshLook24}. 

The upper half of the figure illustrates an ideal online simulation using the same offline dataset. In this simulation, the model is trained/retrained periodically or is a retrieval model. We assume that the model is lastly updated at time $t_m$, which learns from all the interactions $U \times I$ that occurred before $t_m$. We also assume that the test instance of $u_1$ occurred right after the two preceding interactions, which form the $I_c$ of $u_1$. Then predicting $u_1$'s test instance is through a model trained at $t_m$, and $t_m < t_{x_1}$. The model utilizes $u_1$'s historical interactions $I_u$ and her current session $I_c$ to make the prediction, by utilizing the candidate items accessible at time $t_{x1}$. The color gradient shows the dynamics of the items \eg some items are trending at time $t_{x1}$. The dotted box to the right represents the items that are available in the offline dataset but not accessible to the model at time $t_{x1}$. However, it is computationally expensive to strictly follow this online simulation to have a model retrained at every time point of every test instance. A periodical model retraining with the timeline evaluation scheme could be a possible solution~\cite{ji2023retraining,FreshLook24}. Certainly, more research is required to discover the most effective evaluation schemes that best simulate the online setting for RecSys using offline datasets.

In our earlier work~\cite{YuJasist23}, we argue that it is not merely the problem definition in its formal form, but the dataset and training, define the task that a model aims to solve. Put simply, the way a dataset is used defines what information is made available to a model under training. 
To my understanding, with the common practice in RecSys academic research, a model's prediction relies on the general preferences it has learned from a user, while the user's decision-making process takes into account both her general preferences and her current interests during the current interaction session. Yet, the impact of contextual factors varies across recommendation scenarios. Without providing the relevant contextual information, the model learned from these datasets will face a completely different setting when deployed online. Due to the inadequacy of many existing datasets to capture the essential input for users' decision-making processes,  the application of collaborative filtering in predicting users' general and enduring preferences appears to be the main focus of academic research. This could be a possible reason for the significant diverge between RecSys in academic research and RecSys in industry. On the positive side, there is a promising trend in the availability of RecSys datasets containing more information \eg impressions, than simply user-item interactions~\cite{Mind20,Impression22}.

\section{Conclusion}
\label{sec:CF}
This paper revisits RecSys task formulation from a user perspective, highlighting two main messages. Firstly,  RecSys tasks are inherently application-specific, as factors influencing user decision-making vary across different scenarios. Thus, it is imperative to study application-specific recommendation tasks rather than treating all recommendation tasks in a simplified form.  Secondly, recommender systems are dynamic. While collaborative filtering effectively learns general user preferences, it fails to capture dynamics in the decision-making process. Therefore, a balanced approach considering both aspects leads us to conceptualize RecSys as a ranking problem. With a clearer understanding of the RecSys problem, it is hopeful that more datasets containing necessary details will become available for various recommendation scenarios. Furthermore, a more refined task formulation will directly impact the design of effective evaluations. 

Again, all points discussed in this opinion paper are not new. These points should have already been well considered in practical RecSys applications for long, and some points extensively discussed in earlier literature~\cite{AIM22OfflineChallenge}. However, due to the relatively large number of publications from academia, researchers who are new to RecSys may not have well considered these contexts and simply follow the common practice. As a research field closely tied to real-world applications, it is imperative for us to clearly define research tasks tailored to specific recommendation scenarios, rather than relying on a generic and oversimplified setting. This is particularly important when considering new RecSys settings like conversational recommendation, and sequential recommendation, as well as when we bring in new technologies to RecSys like large language models.  Formulating tasks specific to scenarios would also significantly aid in selecting compatible baselines and establishing evaluation protocols  that best simulate practical conditions. Lastly, the definition of scenario-specific tasks heavily depends on the availability of high-quality datasets from real-world platforms.

\section*{ACKNOWLEDGMENTS}
I would like to thank Chenliang Li, Zhaoqi Zhang, Fajie Yuan, Jie Zou,  Xin Zhao, Yitong Ji, Jiayu Li, and Yu-chen Fan for reviewing the draft and providing invaluable comments. Your thoughtful insights and constructive suggestions have contributed much to this manuscript.

\bibliographystyle{sigwebnewsletter} 
\bibliography{recsystask}

\begin{biography}
Dr. Aixin Sun is an Associate Professor at the College of Computing and Data Science, Nanyang Technological University (NTU), Singapore. His research interests include information retrieval, recommender systems,  and social computing. Dr. Sun is an associate editor of ACM TOIS, ACM TALLIP, Neurocomputing, and a member of the editorial board of JASIST. He has also served as Track Chair, Area Chair, Senior PC member, or PC member for several conferences including SIGIR, WWW, WSDM, NeurIPS, and RecSys.
\end{biography}

\end{document}